\documentclass{article}
\usepackage{amssymb}
\title{A Simple Algorithm for Local Conversion of Pure States}
\author{Jens G Jensen and R\"udiger Schack\\
Dept.\ of Mathematics, Royal Holloway, University of London\\
Egham, Surrey TW20 0EX, UK}
\newcommand\ket[1]{\vert#1\rangle}
\begin{document}
\maketitle \abstract{We describe an algorithm for converting one
  bipartite quantum state into another using only local operations and
  classical communication, which is much simpler than the original
  algorithm given by Nielsen [Phys.\ Rev.\ Lett.\ \textbf{83}, 436
  (1999)].  Our algorithm uses only a single measurement by one of the
  parties, followed by local unitary operations which are permutations
  in the local Schmidt bases.}

\section{Introduction}
Consider the case where two parties, Alice and Bob, share an entangled
state of two $N$-level particles.  M.~A.~Nielsen introduced in \cite{N}
an algorithm for converting one such pure bipartite state into another
using only local operations and classical communication.  He gave an
explicit condition for when such a conversion is possible: We write the
first state in Schmidt form
\begin{equation}
\ket\phi=\sum_{i=1}^N\sqrt{\alpha^i}\ket{i}\otimes\ket{i},
\end{equation}
where the $\alpha^i$ are non-negative real numbers satisfying
$\sum_i\alpha^i=1$, and similarly for the target state
\begin{equation}
  \label{eq:psi}
  \ket\psi=\sum_{i=1}^N\sqrt{\beta^i}\ket{i}\otimes\ket{i}.
\end{equation}
Then $\ket{\phi}$ may be converted to $\ket{\psi}$ iff the vector
$(\beta^i)$ \textit{majorizes} $(\alpha^i)$, i.e.,
\begin{equation}
\label{eq:maj}
\forall k:\quad\sum_{i=1}^k{}^\downarrow\beta^i\geq\sum_{i=1}^k{}^\downarrow\alpha^i
\end{equation}
with equality for $k=N$; here $^\downarrow\beta^i$ are the $\beta^i$
arranged in descending order, and similarly for the
$^\downarrow\alpha^i$.  This is again equivalent (theorem II.1.10 of
\cite{Bh}) to the existence of a doubly stochastic matrix $D$ such that
$\alpha=D\beta$ (a doubly stochastic matrix has non-negative real
entries, and all its rows and columns sum to one).

Nielsen's algorithm uses several rounds of individual measurements and
classical communication.  Although it is known that such a sequence of
operations can be replaced by one involving only a single measurement
\cite{LP}, the proof of this is non-constructive and not easily applied to
Nielsen's algorithm. Our simpler algorithm may be
useful for practical applications and for the analysis of local pure-state
conversion in quantum cryptography \cite{B,JS}. A similar result using
a different method has been obtained earlier by Hardy \cite{H}.

\section{Example}

In this section we illustrate how the algorithm works by an example.  We
consider the case $\beta^T=(\beta^i)^T=(3/5,3/10,1/10)$ and
$\alpha^T=(\alpha^i)^T=(2/5,1/4,7/20)$ (for typographic reasons, we
write the transpose of the column vectors); note that $\alpha$ is not
sorted: as we shall see, this doesn't matter.  We check that
$\beta\succ\alpha$.  Using the algorithm from (the proof of) theorem
II.1.10 in \cite{Bh}, we find a doubly stochastic matrix that maps
$\beta$ to $\alpha$:
\begin{equation}
\label{eq:exdsm}
\pmatrix{1/3 & 2/3 & 0 \cr
        1/6 & 1/3 & 1/2 \cr
        1/2 & 0 & 1/2}
\pmatrix{3/5 \cr 3/10 \cr 1/10}
=
\pmatrix{2/5 \cr 1/4 \cr 7/20}
\end{equation}

From this matrix we now derive the set of unitary transformations, which
turn out to be permutations.  We start by finding a set of {\em
  non-zero} entries with no two entries in the same row or in the same
column, i.e., corresponding to some permutation matrix.  We first choose
positions $\pmatrix{0&1&0\cr1&0&0\cr0&0&1}$, corresponding to the
permutation $(12)$.  The smallest entry in the doubly stochastic matrix
in these positions is $1/6$, so we subtract $1/6$ times the permutation
matrix.  Then we get
\begin{equation}
\label{hdahdkhdla}
\pmatrix{1/3 & 1/2 & 0 \cr 0 & 1/3 & 1/2 \cr
1/2 & 0 & 1/3}
\end{equation}
Again we look for a set of non-zero entries with no two in the same row
or column; let's pick the identity matrix this time.  The smallest entry
on the main diagonal of (\ref{hdahdkhdla}) is $1/3$, so we subtract
$1/3$ times the identity matrix, and finally we are left with $1/2$
times a permutation matrix corresponding to the permutation $(132)$.

Using the above decomposition, we can now rewrite (\ref{eq:exdsm}) as
\begin{equation}
\label{eq:exmat}
\pmatrix{ 3/10 & 3/5 & 3/10 \cr 3/5 & 3/10 & 1/10 \cr 1/10 & 1/10 & 3/5 }
\pmatrix{1/6 \cr 1/3 \cr 1/2}
=
\pmatrix{ 2/5 \cr 1/4 \cr 7/20}
\end{equation}
We may imagine that the columns in the matrix are indexed by the
permutations we found above, i.e., $(12)$, $()$, and $(132)$,
respectively.  The entries of column $\sigma$ are then $\beta$ permuted
by $\sigma^{-1}$; we permute the vector by shuffling the components:
$(\beta^1,\beta^2,\beta^3).(123)=(\beta^2,\beta^3,\beta^1)$ (this
corresponds to the last row because $(132)^{-1}=(123)$).  Thus, in the
column corresponding to $\sigma$, the $i$th row has the entry
$\beta^{\sigma^{-1}(i)}$.  The LHS vector in (\ref{eq:exmat}) is
$(p^\sigma)$, where the $p^\sigma$ is the weight corresponding to
$\sigma$ that we found above when decomposing the doubly stochastic
matrix.  The RHS vector is, of course, $\alpha$.

Now we need to find the POVM.  To do this, we find three diagonal
matrices $A_\sigma$, each defined by taking column $\sigma$ and
dividing the $i$th entry (in the $i$th row) with $\alpha^i$, and
multiplying with $p^\sigma$.  Thus
\begin{eqnarray}
A_{(12)}&=\frac16\mathrm{diag}(\frac{3/10}{2/5},\frac{3/5}{1/4},\frac{1/10}{7/20})
        &=\mathrm{diag}(1/8,2/5,1/21) \\
A_{()}&=\frac13\mathrm{diag}(\frac{3/5}{2/5},\frac{3/10}{1/4},\frac{1/10}{7/20})
        &=\mathrm{diag}(1/2,2/5,2/21) \\
A_{(123)}&=\frac12\mathrm{diag}(\frac{3/10}{2/5},\frac{1/10}{1/4},\frac{3/5}{7/20})
        &=\mathrm{diag}(3/8,1/5,6/7)
\end{eqnarray}
We observe that each $A_\sigma$ is a positive matrix, and that the sum 
of the $A_\sigma$ is the identity matrix.  The $A_\sigma$ thus define
a POVM.  All we now need to see is what happens when the corresponding
operations are applied to the state described by $\alpha$.

The POVM is performed locally by one of the parties, say Alice.  In our
example, the post-measurement state corresponding to the outcome
$\sigma$ is a pure state with Schmidt coefficients obtained by
normalizing the vector $A_\sigma\alpha$:
\begin{itemize}
\item $(12)$: $(1/20,1/10,1/60)\rightsquigarrow(3/10,3/5,1/10)$
\item $()$: $(1/5,1/10,1/30)\rightsquigarrow(3/5,3/10,1/10)$
\item $(123)$: $(3/20,1/20,3/10)\rightsquigarrow(3/10,1/10,3/5)$
\end{itemize}
where the $\rightsquigarrow$ indicates the normalization.  The
post-measurement states are thus given by the columns of the matrix in
equation (\ref{eq:exmat}).  To reach the target state, Alice permutes
the bases according to the permutation $\sigma$, and she communicates
$\sigma$ to Bob who then performs the same permutation on his basis.

\section{Formal Description and Proof}

\subsection{Description of the algorithm}

\noindent\textbf{Find the doubly stochastic matrix}

Given vectors of Schmidt coefficients $\alpha$ and $\beta$ such that
$\beta\succ\alpha$, we first use the algorithm of \cite{Bh}, II.1.10,
to find a doubly stochastic matrix $D$ such that $\alpha=D\beta$.

\noindent\textbf{Decompose the doubly stochastic matrix}

The Birkhoff-von Neumann theorem (\cite{Bh}, chapter 2) states that we
can find permutations $\sigma\in\Sigma\subseteq\mathfrak{S}_N$ and
positive numbers $p^\sigma$ such that
\begin{equation}
\label{eq:decomp}
D=\sum_{\sigma\in\Sigma}p^\sigma P_\sigma
\end{equation}
where $P_\sigma$ is the permutation matrix associated to $\sigma$,
i.e., it maps the fundamental vector $e_i$ (with a one in the $i$th
place and zeros otherwise) to $e_{\sigma(i)}$.

To explicitly find the decomposition of the density matrix, we may use
that many of the proofs of the Birkhoff-von Neumann theorem are
constructive.  The approach we sketch here is similar to the proof in
\cite{Pul}.

It will be useful to define an \textit{arrangement} (or
\textit{diagonal}) of a $n\times n$ matrix $M=(m^i_j)$ as a set of
entries $\{m_i^{\sigma(i)}\}$ for some permutation
$\sigma\in\mathfrak{S}_n$, i.e., where no two entries are taken from the
same row or the same column.  It follows from the
K\"onig-Frobenius theorem (\cite{Bh}, chapter~2) that any doubly
stochastic matrix has an arrangement with all entries non-zero.

There is a polynomial-time algorithm for finding such arrangements in a
matrix: Consider the matrix $\tilde{D}$ which is $D$ with each non-zero
entry replaced by $1$.  Let $R=\{r_1,\dots,r_n\}$ be the set of row
indices of $\tilde{D}$ and $C=\{c_1,\dots,c_n\}$ the set of column
indices.  Then $\tilde{D}$ defines a balanced bipartite graph $G$ with
vertices $R\cup C$ (disjoint union): we have an edge from $r_i$ to $c_j$
iff $\tilde{D}^i_j=1$.  By the K\"onig-Frobenius theorem, $G$ has no
isolated nodes.  We can thus easily find a perfect matching in the
graph; there are polynomial-time algorithms for doing this, see e.g.,
chapter~3, section~3 of \cite{Pul}.  A perfect matching defines a unique
permutation $\sigma\in\mathfrak{S}_n$ through its edges
$r_{\sigma(i)}\leftrightarrow c_i$, $i=1,\dots,n$.  It is easy to see
that the matching also defines an arrangement of $D$ with all entries
non-zero.

\noindent\textbf{Constructing the POVM}
\label{sec:povm}

Now we assume that we have the decomposition (\ref{eq:decomp}).  Hence
\begin{equation}
\label{eq:crux}
\forall i:\quad\sum_\sigma\beta^{\sigma^{-1}(i)}p^\sigma=\alpha^i
\end{equation}
Indeed, $\alpha^i=(D\beta)^i=[(\sum_\sigma p^\sigma P_\sigma)\beta]^i=\sum_\sigma p^\sigma\beta^{\sigma^{-1}(i)}$.

Next, we define matrices
\begin{equation}
\label{eq:povm}
A_\sigma=p^\sigma\mathrm{diag}_i({{\beta^{\sigma^{-1}(i)}}\over{\alpha^i}})
\end{equation}
where $\mathrm{diag}_i(f(i))$ denotes a diagonal matrix whose
$(i,i)$ entry is $f(i)$.

The $\{A_\sigma\mid\sigma\in\Sigma\}$ define a POVM, since each
$A_\sigma$ is a positive matrix, and
\begin{eqnarray}
\sum_\sigma A_\sigma&=&\sum_\sigma p^\sigma\mathrm{diag}_i\left({{\beta^{\sigma^{-1}(i)}}\over{\alpha^i}}\right) \nonumber\\
&=&\mathrm{diag}_i\left({1\over{\alpha^i}}\sum_\sigma p^\sigma\beta^{\sigma^{-1}(i)}\right) \nonumber\\
&=&\mathrm{diag}_i1 \nonumber
\end{eqnarray}

Furthermore,
\begin{equation}
  \label{eq:psigma}
  \mathrm{tr}(A_\sigma\mathrm{diag}_i(\alpha^i))=p^\sigma
\end{equation}
so that $p^\sigma$ is the probability of outcome $\sigma$.

\noindent\textbf{The measurement}

Now we turn to the measurement itself.  When Alice performs the
measurement on her side, we need only consider her reduced density
matrix, $\rho$, which has eigenvalues $\alpha^i$.  We choose the
operations corresponding to the POVM $A_\sigma$ such that, if the
outcome of the measurement is $\sigma$, then the reduced density matrix
after the measurement is
\begin{equation}
  \label{eq:dsajhfchgbfdslk}
  \rho_\sigma=\frac1{p^\sigma}\sqrt{A_\sigma}\rho\sqrt{A_\sigma}=\frac1{p^\sigma}A_\sigma\rho.
\end{equation}
The eigenvalues of $\rho_\sigma$ are $\beta^{\sigma^{-1}(i)}$,
$i=1,\dots,N$, which sum to one.  Alice then performs the permutation
$\sigma$ on her side, and communicates $\sigma$ to Bob so that he
can do the same: since
$\sum_i\beta^{\sigma^{-1}(i)}\ket{i}\otimes\ket{i}=\sum_i\beta^i\ket{\sigma(i)}\otimes\ket{\sigma(i)}$,
the operation should map $\ket{\sigma(i)}$ to $\ket{i}$ for all $i$.
After the unitary operations, the particles are in the pure state
$\ket\psi$ (\ref{eq:psi}) with Schmidt coefficients $\beta^i$.

\subsection{The converse}

We now turn to the converse: given $\alpha$ and $\beta$ and a POVM
$\{A_\sigma\mid\sigma\in\Sigma\}$ with $\Sigma\subseteq\mathfrak{S}_n$
such that for all $\sigma\in\Sigma$,
\begin{equation}
  \label{eq:conv}
  (A_\sigma\alpha).\sigma=p^\sigma\beta,
\end{equation}
for some positive constants $p^\sigma$, we show that $\beta\succ\alpha$
(of course, this also follows directly from Nielsen's theorem \cite{N}).
We assume that each $A_\sigma$ is given by a diagonal matrix.

Define a matrix $\Gamma=(\gamma^i_j)$ by
\begin{equation}
  \label{eq:gamma}
  \gamma^i_j=\sum_{\{\sigma\mid\sigma(j)=i\}}p^\sigma
\end{equation}

Then we have
\begin{eqnarray}
  \label{eq:check}
  \forall j:\quad\sum_i\gamma^i_j&=&\sum_i\sum_{\{\sigma\mid\sigma(j)=i\}}p^\sigma \nonumber\\
  &=& \sum_{\sigma\in\Sigma}p^\sigma=1\;; \nonumber\\
  \forall i:\quad\sum_j\gamma^i_j&=&\sum_j\sum_{\{\sigma\mid\sigma(j)=i\}}p^\sigma \nonumber\\
  &=& \sum_{\sigma\in\Sigma}p^\sigma=1\;,\nonumber
\end{eqnarray}
so $\Gamma$ is doubly stochastic.  Now equation (\ref{eq:conv}) implies
$A_\sigma\alpha=p^\sigma\beta.(\sigma^{-1})$, which again implies that
$A_\sigma$ must be of the form (\ref{eq:povm}).  Finally,
\begin{eqnarray}
  \label{eq:dhkahdki}
  \forall i:\quad(\Gamma\beta)^i&=&\sum_j\gamma^i_j\beta^j\nonumber\\
  &=&\sum_j\sum_{\{\sigma\mid\sigma(j)=i\}}p^\sigma\beta^{\sigma^{-1}(i)}\nonumber\\
  &=&\sum_{\sigma\in\Sigma}p^\sigma\beta^{\sigma^{-1}(i)}\nonumber\\
  &=&\alpha^i.
\end{eqnarray}
Thus $\Gamma\beta=\alpha$, which means that $\beta\succ\alpha$.

\section{Acknowledgment}

This work was motivated by a discussion in the Quantum Dynamics group at
Royal Holloway between N.~L\"utkenhaus and the authors.  This work is
supported by the UK Engineering and Physical Sciences Research Council
(EPSRC).

\end{document}